\documentclass[aps,prd,twocolumn,reprint,floatfix,preprintnumbers,nofootinbib]{revtex4-2}

\usepackage{graphicx,ytableau,placeins}
\usepackage{amsmath,amssymb,amsfonts}
\usepackage[caption=false]{subfig}
\usepackage[colorlinks=True, citecolor=blue, urlcolor=blue, linkcolor=blue]{hyperref}
\usepackage{comment}
\usepackage{tikz}
\usetikzlibrary{math}
\usetikzlibrary{calc}

\newcommand{\be}{\begin{equation}}
\newcommand{\ee}{\end{equation}}
\newcommand{\ba}{\begin{eqnarray}}
\newcommand{\ea}{\end{eqnarray}}
\newcommand{\ban}{\begin{eqnarray*}}
\newcommand{\ean}{\end{eqnarray*}}

\usepackage{ulem}

\begin{document}

\title{Gluon Entanglement Entropy Inside a Nucleon: A Toy Model}
\author{David Horn}
\email{david.horn@duke.edu}
\affiliation{Department of Physics, Duke University, Durham, North Carolina 27708, USA}
\author{Berndt M\"uller}
\email{berndt.mueller@duke.edu}
\affiliation{Department of Physics, Duke University, Durham, North Carolina 27708, USA}
\author{Xiaojun Yao}
\email{xjyao@uw.edu}
\affiliation{InQubator for Quantum Simulation, Department of Physics,
University of Washington, Seattle, Washington 98195, USA}
\date{\today}
\preprint{IQuS@UW-21-126}

\begin{abstract}
We construct a toy model of a nucleon, in which three static quarks interact via a SU(3) gauge field on a planar honeycomb lattice. The dynamics of the gauge field is described by the Kogut-Susskind Hamiltonian, truncated to the lowest three SU(3) irreducible representations. We show that the internal structure of the toy nucleon reflects salient features of the physical nucleon state. We then find the entanglement entropy of the gauge field within the nucleon state and compute its time evolution after a quench, in which all three valence quarks are suddenly removed. We show that the entanglement entropy in the final state is dominated by the dynamically generated contribution rather than the initial state entropy.
\end{abstract} 

\maketitle

At a fundamental level, all high-energy reactions involving elementary particles are governed by a unitary S-matrix.
However, as it is impossible to experimentally measure all details of the final state in a reaction that produces many particles, the coarse-grained final state is often described as a sample from a statistical particle distribution that carries a non-vanishing entropy.
This interpretation dates back to Fermi's statistical model \cite{Fermi:1950jd} of multiparticle production, which posits that the entropy of the final particle distribution is microcanonically maximal.
At its most inclusive level, the statistical model has been found to give a good description of hadron yields in reactions ranging from heavy-ion collisions to electron-positron annihilation \cite{Cleymans:1992zc,Becattini:2008tx}. 

The origin of this coarse-grained entropy of the multiparticle final states has long been debated.
Various arguments have invoked the decoherence of a high-density semi-classical gluon state in nuclear collisions \cite{Kovner:1995ja,Fries:2008vp}, the QCD analogue of Hawking-Unruh radiation \cite{Castorina:2007eb}, and the pre-formation of the entanglement entropy of partons inside the participating hadrons \cite{Kharzeev:2017qzs,Baker:2017wtt,Tu:2019ouv,Kharzeev:2021yyf,Zhang:2021hra,Gursoy:2023hge}, which is released by the reaction.
In particular, it has been argued that the coarse-grained entropy characterizing the final state of a deep inelastic scattering reaction $e^-+p\to e^-+X$, which is usually obtained from the single-particle momentum distribution, is proportional to the proton's gluon distribution function $G(x,Q^2)$ \cite{Kharzeev:2017qzs,Kharzeev:2021yyf} and can be interpreted as the entanglement entropy of low-$x$ gluons contained in the proton's wave function.
On the other hand, in the QCD-string breaking model of multi-particle production in high-energy $e^+e^-\to q\overline{q}$ annihilation, the flux-tube between the quark-antiquark pair is modeled as a coherent gluon  field, and the final-state entropy is generated by string breaking \cite{Florio:2023dke,Grieninger:2023ufa,Lee:2023urk,Florio:2024aix,Liu:2024lut,Ikeda:2025bjb,Florio:2025gic,Florio:2025hoc,Grieninger:2026bdq}.

In a deep inelastic scattering reaction, a single quark is suddenly ejected from the ground state nucleon after it has absorbed a high-energy, highly virtual photon.
This process cannot be easily modeled on a small lattice, but we can create a setup that allows us to study the consequences of an even more drastic process, the simultaneous removal of {\it all three} color charges representing the valence quarks.
This process, which cannot be easily realized experimentally, can be thought of as an idealized version of a high-energy reaction, where three valence quarks are suddenly ejected from the nucleon rather than a single one.
Using a simplified model of a nucleon as a bound state of three static quarks, we compute the entanglement entropy of a localized region of gauge field in the toy nucleon ground state and trace its time evolution after the quench-like removal of the three color charges. 

\begin{figure}[htb]
  \includegraphics[width=0.75\linewidth]{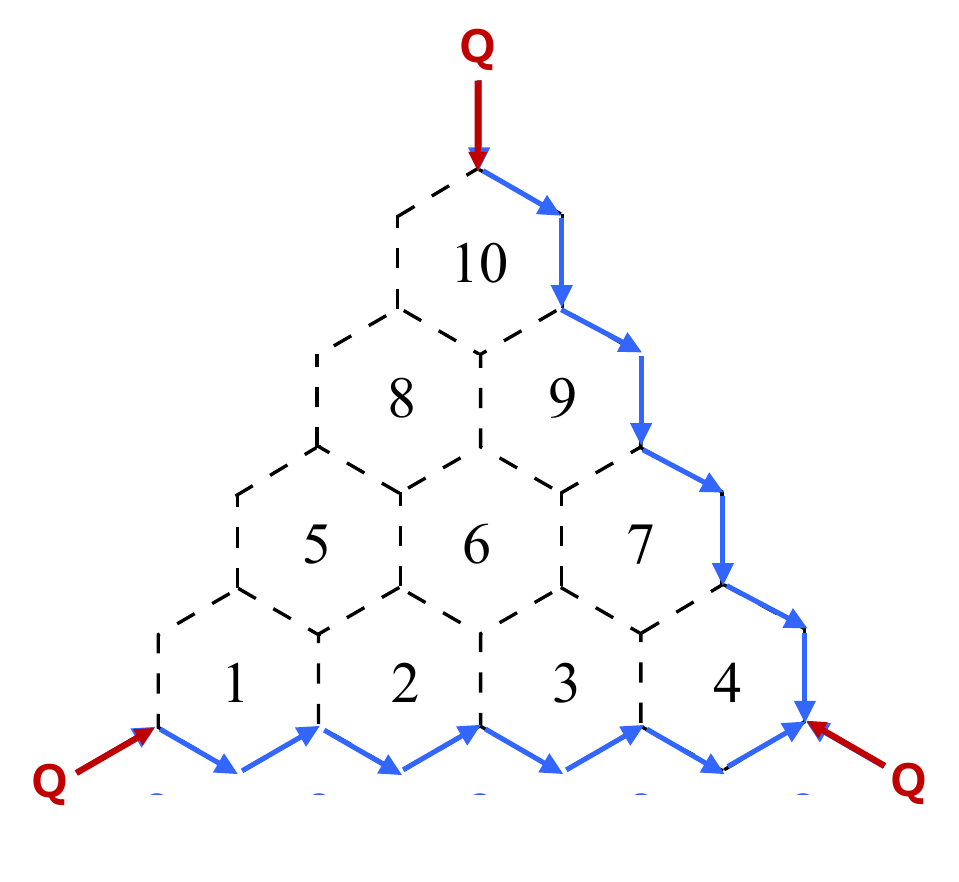}
\caption{Triangular lattice configuration with three static fundamental SU(3) charges at its corners. The colored lines with arrows indicate the chromoelectric flux boundary condition that ensures that the overall configuration is a color singlet. We consider triangular lattices with 10 (shown here) and 15 plaquettes.}
\label{fig:QQQ}
\end{figure}

{\it The toy model.} We now construct a numerically exactly solvable toy model of a nucleon that allows us to explore the relative importance of initial-state entanglement entropy residing in the gluon field inside a nucleon and the entropy production by decoherence among the different components of the excited nucleon state after the onset of the reaction. In our setup a fast-moving nucleon is modeled by three static fundamental SU(3) color charges (the valence quarks) attached to the corners of a two-dimensional triangular lattice as shown in Fig.~\ref{fig:QQQ}. One can think of the planar lattice as a slice of the nucleon state at a given rapidity, where the valence quarks appear frozen because of time dilation. 
The dynamical gauge field is described by the Kogut-Susskind Hamiltonian in the minimal truncation of the SU(3) gauge group, which includes the irreps (0,0), (1,0) and (0,1) in Dynkin notation. A similar configuration has recently been used to compute the string tension associated with three static color charges from the energy contained in the gauge field on a hexagonal $4\times 4$ lattice \cite{Chen:2026hnh}.
Here we choose a triangular lattice to better represent the geometrical symmetry in our toy nucleon.

The SU(3) Kogut-Susskind Hamiltonian for a planar hexagonal lattice is given by
\be
H = \frac{g^2\Sigma}{3}\sum_{\rm links} (E_\ell^a)^2 - \frac{1}{g^2\Sigma}\sum_{\rm plaq}(U_P+U_P^\dagger)\,,
\label{eq:HKS}
\ee
where $g$ is the gauge coupling, $\Sigma=3\sqrt{3}/2$, and $U_P$ denotes the plaquette operator.
We denote the state of the plaquette in the qutrit representation: $|0\rangle = (1,0,0)$ corresponds to the electric flux ground state.
Applying the plaquette operator gives $U_P |0\rangle \equiv_3 |1\rangle = (0,1,0)$, $U_P |1\rangle \equiv_3 |2\rangle = (0,0,1)$, and $U_P |2\rangle \equiv_3 |0\rangle$, where $\equiv_3$ denotes the {\it modulo} 3 operation.
For details of the calculation of the Hamiltonian matrix elements in the qutrit basis we refer to Chen {\it et al.} \cite{Chen:2026hnh}. 
Here we consider gauge couplings $g^2=0.3$ and $g^2=0.5$, which lie in the ergodic domain for this lattice Hamiltonian.
The Hilbert space on which the Hamiltonian \eqref{eq:HKS} acts depends on the state of the external links of the lattice, which represent external static charges and serve as fixed boundary conditions.
Only the internal and edge links and plaquettes, shown as dashed lines or blue lines with arrowheads in Fig.~\ref{fig:QQQ}, are dynamical; the brown links injecting chromoelectric flux at the triangle corners are static.

The ground state of the lattice Hamiltonian accordingly depends on the external color charges.
We denote the vacuum, i.e., the ground state in the absence of external charges (all external links are in the singlet representation), as $|{\rm vac}\rangle$, and the ground state in the presence of three external color charges in the fundamental representation at the locations indicated in Fig.~\ref{fig:QQQ} (only external links in the location of the static quarks are chosen in the triplet representation) as $|\textrm{QQQ}\rangle$.
This state serves as our model for the fast moving nucleon. 

Before we compute the entanglement entropy of the gauge field and its time evolution after the removal of the quarks, we explore the energy density and planar pressure {\it per plaquette} inside our toy nucleon.
These can be obtained from the stress-energy tensor and written as
\begin{align}
\epsilon &= \frac{g^2\Sigma}{3} \sum_{\ell \in \partial P} c_\ell (E^a_\ell)^2 - \frac{1}{g^2\Sigma}(U_P+U_P^\dagger) \,,\\
P_\mathrm{pl} &= \frac{P_{xx} + P_{yy}}{2} = \epsilon -  \frac{g^2\Sigma}{3} \sum_{\ell \in \partial P} c_\ell (E^a_\ell)^2 \,,
\end{align}
where $P$ denotes a plaquette and the sum is over all links $\ell$ forming the perimeter $\partial P$ of the plaquette. The factor $c_\ell$ differs for contributions from edge links and internal links, the latter being shared between two neighboring plaquettes. Specifically, we assign $c_\ell=1$ to edge links and $c_\ell=1/2$ to internal links.

\begin{figure}
    \centering
    \includegraphics[width=1\linewidth]{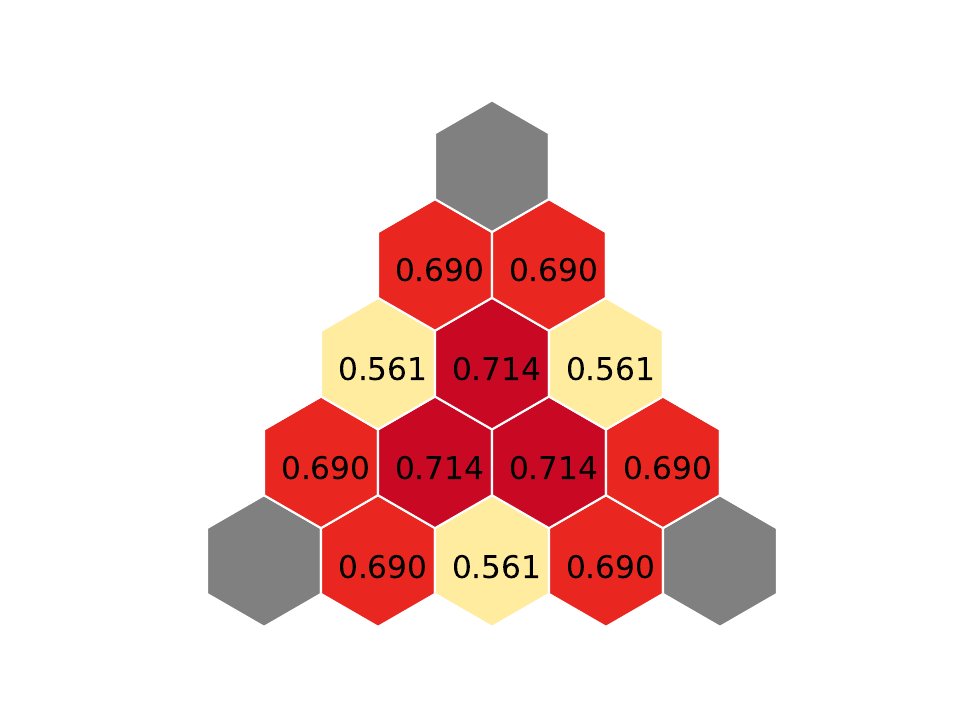}
    \caption{Vacuum subtracted energy densities of the gauge field per plaquette for the $N=5$ lattice at a coupling of $g^2=0.5$. Dark red: body-centered plaquettes; pale yellow: edge-centered plaquettes, light red: off-center plaquettes.}
    \label{fig:energy densities}
\end{figure}

\begin{figure}
    \centering
    \includegraphics[width=1\linewidth]{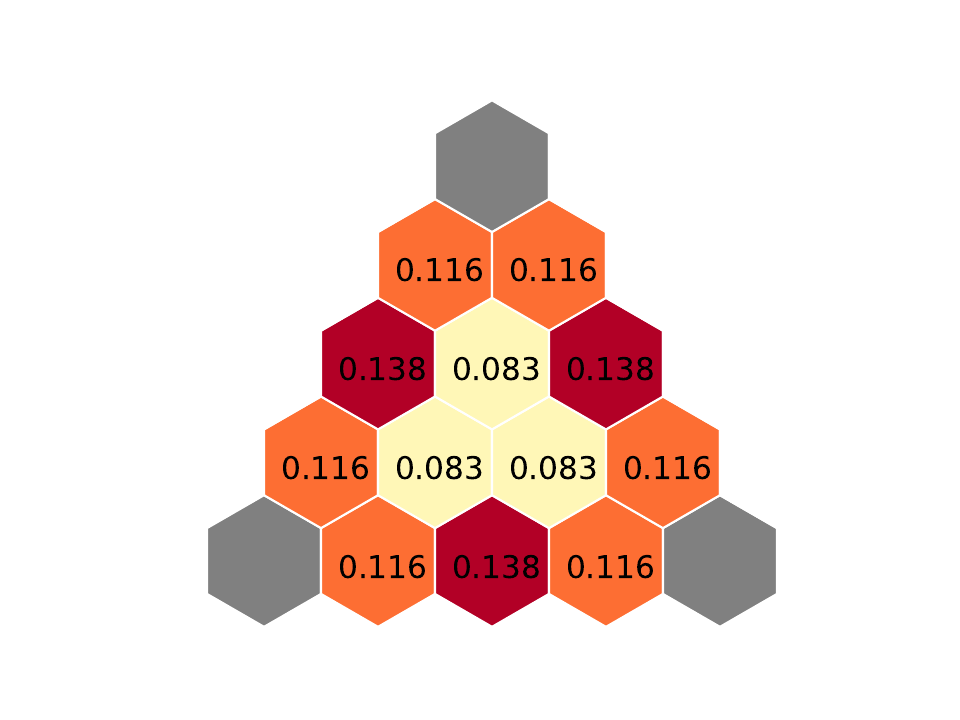}
    \caption{Vacuum subtracted planar pressure of the gauge field per plaquette for the $N=5$ lattice at a coupling of $g^2=0.5$. Note the different color coding -- Pale yellow: body-centered, dark red: edge-centered, orange: off-center.}
    \label{fig:planar pressure}
\end{figure}

The (vacuum subtracted) energy densities and planar pressures for the $N=5$ lattice are given in Figs.~\ref{fig:energy densities} and \ref{fig:planar pressure} for a coupling of $g^2 = 0.5$. We use color coding to indicate the magnitude of the energy density and pressure for the different types of plaquettes, with darker colors indicating larger values.
Table~\ref{tab:energy densities and pressures} shows these values for $g^2 = 0.5$ and a weaker coupling $g^2 = 0.3$ for the three different types of plaquettes that exist in the $N=5$ lattice (body-centered, edge-centered and off-center) labeled by color in Figs.~\ref{fig:energy densities} and \ref{fig:planar pressure}.
As Table~\ref{tab:energy densities and pressures} shows, the energy density distribution inside the nucleon undergoes a structural change between the two couplings.
At the larger coupling ($g^2=0.5$) the energy density reflects a Y-shaped, junction-like configuration shown in Fig~\ref{fig:energy densities}, whereas at the lower coupling the gauge field configuration changes to a triangular shape. 
It is known from quenched lattice-QCD simulations \cite{Bissey:2005sk} that the gauge field distribution in the nucleon exhibits such a junction shape when the valence quarks are well separated (see Fig.~6 in \cite{Bissey:2005sk}) and the running QCD coupling is large. 

The pressure distribution inside the proton is experimentally known to be positive (repulsive) at short radii and negative (attractive) at large radii \cite{Burkert:2018bqq}. A theoretical analysis \cite{Fujii:2025aip} shows that the dynamical pressure component is positive everywhere (see Fig.~2 in \cite{Fujii:2025aip}). The pressure component due to the trace anomaly, which is absent in our toy model, is negative and dominates at large radii.
\begin{table}[]
    \centering
    \begin{tabular}{|c|c c | c c|}
    \hline
         & \multicolumn{2}{c|}{$\langle \epsilon \rangle - \langle\epsilon\rangle_\mathrm{vac}$} & \multicolumn{2}{c|}{$\langle P_\mathrm{pl}\rangle - \langle P_\mathrm{pl}\rangle_\mathrm{vac}$} \\
         \hline
        $g^2$ & 0.3 & 0.5 & 0.3 & 0.5 \\
         \hline
        body-centered & 0.425 & 0.714 & 0.166 & 0.083 \\
        edge-centered & 0.619 & 0.561 & 0.334 & 0.138 \\
        off-center & 0.493 & 0.690 & 0.143 & 0.116 \\
        \hline
    \end{tabular}
    \caption{Vacuum subtracted energy density and planar pressure values for the $N=5$ lattice at the couplings $g^2=0.5$ and $g^2=0.3$. For the definition of the different plaquette types, see Figs.~\ref{fig:energy densities} and \ref{fig:planar pressure}.}
    \label{tab:energy densities and pressures}
\end{table}

Before computing the entanglement entropy of the gauge field, we list some of the shortcomings of our toy model: 
(1) Our valence quarks are static and not dynamical and thus do not carry entropy themselves. 
(2) Our toy nucleon is two-dimensional, not three-dimensional. 
(3) Our nucleon is not moving so it has much lower energy and thus potentially lower entanglement entropy than a fast-moving nucleon. 
(4) As level densities grow rapidly with the space  dimension and local Hilbert space size (such as the number of irreps included in the basis) of a quantum system, our model can be expected to underestimate the absolute magnitude of the entanglement entropies of the initial and final states. 
(5) Being constrained by the fixed lattice, the excitation energy created by the sudden removal of the quarks cannot disperse in space. These deficiencies can only be removed by a full dynamical quantum calculation of QCD with dynamical quarks in (3+1) dimensions, which is currently impossible.

{\it Entanglement entropy.} We now turn to discuss the entanglement entropy of gluons inside our toy nucleon and its time dependence. 
Although the toy nucleon represents a pure quantum state, the gauge field on a geometrically restricted subregion $A$ of the lattice is given by a density matrix:
\ba
\rho_A^\textrm{vac} &=& {\rm Tr}_{\bar{A}}\, |{\rm vac}\rangle\langle{\rm vac}| \
\\
\rho_A^{QQQ} &=& {\rm Tr}_{\bar{A}}\, |\textrm{QQQ}\rangle\langle\textrm{QQQ}|,
\ea
where $\bar{A}$ denotes the complement of the region $A$.
As is well known, even the vacuum state has a nonzero entanglement entropy $S_A^\textrm{vac} = - {\rm Tr}_A \left(\rho_A^\textrm{vac}\ln\rho_A^\textrm{vac}\right)$ when restricted to a finite region $A$.
For theories with a mass gap, $S_A^\textrm{vac}$ is proportional to the surface area of the domain $A$.
The vacuum subtracted entanglement entropy of the gauge field in our toy nucleon state for the domain $A$ is given by
\be
S_A^\textrm{nuc} = -{\rm Tr}_A \left(\rho_A^{QQQ} \ln\rho_A^{QQQ} \right) - S_A^\textrm{vac}.
\ee
We interpret this quantity as the entanglement entropy of the gauge field with respect to a localized domain within our toy nucleon.

When we suddenly remove the external charges by setting all external links to the singlet state, the nucleon state $|\textrm{QQQ}\rangle$ is no longer an eigenstate of the Hamiltonian, but corresponds to a coherent superposition of excited states of the SU(3) vacuum.
We can track the time evolution of this state by evolving the gauge field configuration of the $|\textrm{QQQ}\rangle$ state in the eigenbasis $|\alpha\rangle$ of the vacuum sector of the Hamiltonian \eqref{eq:HKS}:
\be
|\psi(t)\rangle = \sum_\alpha e^{-iE_\alpha t} |\alpha\rangle\langle\alpha|\textrm{QQQ}\rangle
\label{eq:psit}
\ee
with the initial condition $|\psi(0)\rangle=|\textrm{QQQ}\rangle$. The entanglement entropy of the state $|\psi(t)\rangle$ is denoted as $S_A^{QQQ}(t)$.

\begin{figure}[htb]
\centering  \includegraphics[width=0.95\linewidth]{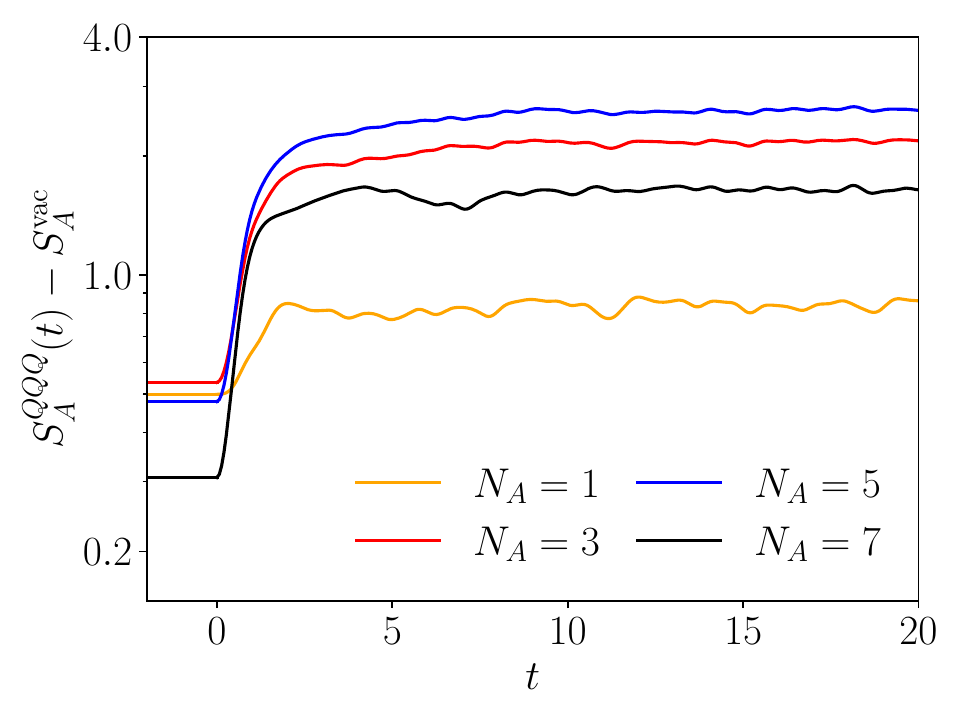}
\includegraphics[width=0.95\linewidth]{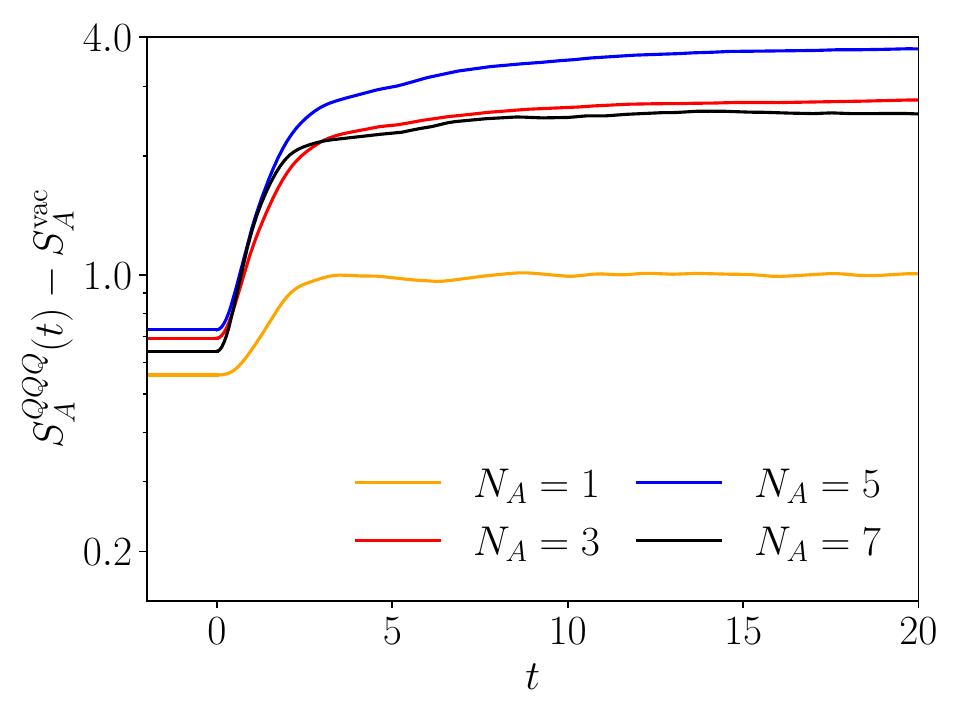}
\caption{Vacuum subtracted entanglement entropies $\Delta S_A(t) = S^{QQQ}_A(t) - S_A^{\rm vac}$ of subsystems of size $N_A=1,3,5,7$ of the toy nucleon as a function of time. The top panel is for $g^2=0.3$; the bottom panel is for $g^2=0.5$. The values for $t\leq 0$ represent the gluon entanglement entropies in the original toy nucleon ground state. Note that the entropy is shown on a logarithmic scale.}
\label{fig:SAQQQt}
\end{figure}

We are now ready to discuss the time evolution of the gluon entanglement entropy in our toy nucleon, beginning with the $N=4$ lattice shown in Fig.~\ref{fig:QQQ}.
Figure~\ref{fig:SAQQQt} shows the vacuum subtracted entanglement entropy $\Delta S_A(t) = S^{QQQ}_A(t) - S_A^{\rm vac}$ for four different subdomains of the toy nucleon shown in Fig.~\ref{fig:QQQ} comprising $N_A = 1,3,5,7$ plaquettes, respectively.
Using the labeling of the plaquettes in Fig.~\ref{fig:QQQ}, these are given by the subregions $\{6\}$, $\{2,3,6\}$, $\{2,3,5,6,7\}$, and $\{2,3,5,6,7,8,9\}$.
For those regions that are not symmetric under $60^\circ$ rotations, we average our results over the plaquettes obtained by rotation by multiples of $60^\circ$.

We see that only a fraction of the final equilibrated entropy can be attributed to the entanglement entropy of the gauge field in the toy nucleon, shown by the value of $\Delta S_A(t)$ for $t\leq 0$.
Numerical values are shown in Table~\ref{EEQQQfrac}.
We also note that the relative initial entropy increase is faster for $g^2=0.3$, but the initial and final values of the  entanglement entropy are larger for $g^2=0.5$ due to the increased energy stored in the gauge field for the larger coupling.

\begin{table}[htb]
\centering
\begin{tabular}{|c|c|c|c|c|c|c|c|c|}
\hline
$g^2$ & \multicolumn{4}{|c|}{0.3} & \multicolumn{4}{|c|}{0.5} \\
\hline
$N_A$ & 1 & 3 & 5 & 7 & 1 & 3 & 5 & 7 \\
\hline
$\Delta S_A(0)$ & 0.499 & 0.535 & 0.479 & 0.308 & 0.558 & 0.692 & 0.727 & 0.642 \\
$\Delta S_A(20)$ & 0.861 & 2.184 & 2.607 & 1.642 & 1.007 & 2.770 & 3.732 & 2.554 \\
$S_A^{\rm vac}$ & 0.064 & 0.140 & 0.175 & 0.169 & 0.026 & 0.048 & 0.056 & 0.050 \\
\hline
\end{tabular}
\caption{Vacuum subtracted entanglement entropies $\Delta S_A$ for different subregions of the toy nucleon in Fig.~\ref{fig:QQQ} for $g^2=0.3$ and $0.5$ at $t=0$ and $t=20$. The entanglement entropies of the subregions in the vacuum state are also shown. As can be seen, only a small fraction of the final entropy can be attributed to the entanglement entropy of the gauge field inside the unperturbed nucleon state.}
\label{EEQQQfrac}
\end{table}

\begin{figure}[t]
\centering  \includegraphics[width=0.95\linewidth]{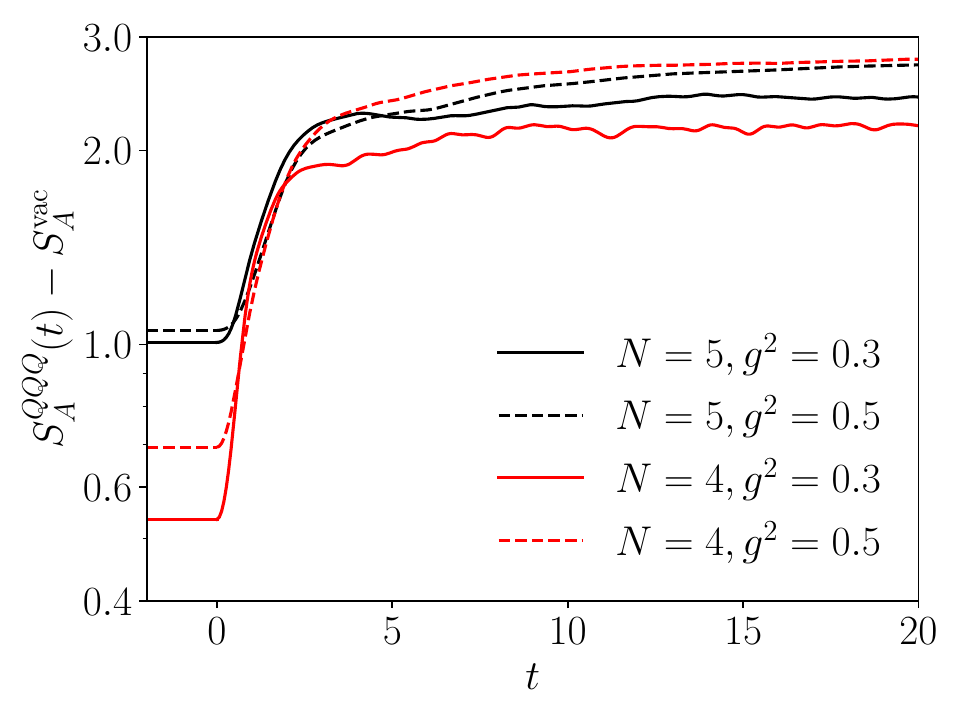}
\caption{Comparison of the vacuum subtracted entanglement entropies $\Delta S_A$ of a three-plaquette subsystem ($N_A=3$) for two different coupling constants $g^2=0.3$ (solid curves) and $g^2=0.5$ (dashed curves) and two lattice sizes $N=4$ (red curves) and $N=5$ (black curves), where $N$ indicates the side length of the triangular lattice. The subsystem is the same as shown in Fig.~\ref{fig:SAQQQt} for $N=4$ and corresponds to the central three plaquettes that do not overlap with the edges for $N=5$. The values for $t\leq 0$ represent the gluon entanglement entropies in the original toy nucleon ground state. Note that the entropy is shown on a logarithmic scale.
}
\label{fig:SAQQQt_comp}
\end{figure}

In Fig.~\ref{fig:SAQQQt_comp} we compare the vacuum subtracted entanglement entropies $\Delta S_A$ of a three-plaquette subsystem ($N_A=3$) for two lattice sizes $N$ and two different coupling constants $g^2 = 0.3$ (solid curves) and $g^2 = 0.5$ (dashed curves).
Here $N$ denotes the side length of the triangular lattice.
In particular, $N=4$ lines (red) refer to the 10-plaquette lattice shown in Fig.~\ref{fig:QQQ} while $N=5$ lines (black) indicate a triangular lattice composed of 15 hexagonal plaquettes.
Unsurprisingly, the initial-state entanglement entropy is larger for $N=5$ than $N=4$ because the gauge field configuration in the probe region $N_A=3$ is less constrained by the requirement that the overall field configuration is a pure quantum state.

{\it Conclusions.} We have constructed a simplified model of a nucleon with three static quarks that allows us to numerically compute the entanglement entropy of the gauge field in the interior of the nucleon and its time evolution after a sudden perturbation. The gauge field binding the quarks has features that mirror those known from Euclidean lattice QCD simulations and experiment. We have found that the final-state entropy in the gauge field after the sudden removal of the valence quarks far exceeds the entanglement entropy of the gauge field in the unperturbed nucleon state. We conclude that attempts to ``measure'' the gluon entanglement entropy in deep-inelastic scattering, where one quark is ejected violently from the nucleon, suffer from severe contamination by the statistical entropy created in the far out-of-equilibrium evolution of the final state.

Finally, while it is possible to compute the entanglement entropy of a region within the nucleon using the tools of Euclidean lattice gauge theory \cite{Buividovich:2008kq,Itou:2015cyu,Rabenstein:2018bri}, it is more challenging to measure this entropy experimentally. As our model study has shown, measurements of entanglement entropy based on processes that destroy the nucleon, such as inelastic scattering, are at least contaminated, if not dominated by statistical entropy created by the measurement process. It is an open question whether experimental processes using nondestructive measurements can be devised that can reliably determine the entanglement entropy of regions within the nucleon ground state.

\bigskip
{\it Acknowledgments:}
     Our study was motivated by remarks by K.~Rajagopal in a lecture at the Kavli Institute for Theoretical Physics (\url{https://online.kitp.ucsb.edu/online/quarkgluon-c26/rajagopal/}). We thank D.~Kharzeev and K.~Rajagopal for comments on a near-final version of the manuscript. 
     D.H. and B.M. are supported by the National Science Foundation (Project PHY-2434506). B.M. also acknowledges support by the U.S. Department of Energy, Office of Science (Grant DE-FG02-05ER41367). X.Y. is supported by the U.S. Department of Energy, Office of Science, Office of Nuclear Physics, InQubator for Quantum Simulation (IQuS) under DOE (NP) Award Number DE-SC0020970 via the program on Quantum Horizons: QIS Research and Innovation for Nuclear Science. This research used resources of the National Energy Research Scientific Computing Center (NERSC), a Department of Energy Office of Science User Facility using NERSC award NP-ERCAP0032083. This work was enabled, in part, by the use of advanced computational, storage and networking infrastructure provided by the Hyak supercomputer system at the University of Washington.
     This research was supported in part by grant NSF PHY-2309135 to the Kavli Institute for Theoretical Physics (KITP).

\bibliographystyle{apsrev4-1}
\bibliography{GluonEE}

@article{Castorina:2007eb,
	archiveprefix = {arXiv},
	author = {Castorina, P. and Kharzeev, D. and Satz, H.},
	doi = {10.1140/epjc/s10052-007-0368-6},
	eprint = {0704.1426},
	journal = {Eur. Phys. J. C},
	pages = {187--201},
	primaryclass = {hep-ph},
	reportnumber = {BNL-NT-07-18, BI-TP-2007-06},
	title = {{Thermal Hadronization and Hawking-Unruh Radiation in QCD}},
	volume = {52},
	year = {2007}}

@article{Tu:2019ouv,
	archiveprefix = {arXiv},
	author = {Tu, Zhoudunming and Kharzeev, Dmitri E. and Ullrich, Thomas},
	doi = {10.1103/PhysRevLett.124.062001},
	eprint = {1904.11974},
	journal = {Phys. Rev. Lett.},
	number = {6},
	pages = {062001},
	primaryclass = {hep-ph},
	title = {{Einstein-Podolsky-Rosen Paradox and Quantum Entanglement at Subnucleonic Scales}},
	volume = {124},
	year = {2020}}

@article{Gursoy:2023hge,
	archiveprefix = {arXiv},
	author = {G{\"u}rsoy, Umut and Kharzeev, Dmitri E. and Pedraza, Juan F.},
	doi = {10.1103/PhysRevD.110.074008},
	eprint = {2306.16145},
	journal = {Phys. Rev. D},
	number = {7},
	pages = {074008},
	primaryclass = {hep-th},
	reportnumber = {IFT-UAM/CSIC-23-79},
	title = {{Universal rapidity scaling of entanglement entropy inside hadrons from conformal invariance}},
	volume = {110},
	year = {2024}}

@article{Zhang:2021hra,
	archiveprefix = {arXiv},
	author = {Zhang, Kun and Hao, Kun and Kharzeev, Dmitri and Korepin, Vladimir},
	doi = {10.1103/PhysRevD.105.014002},
	eprint = {2110.04881},
	journal = {Phys. Rev. D},
	number = {1},
	pages = {014002},
	primaryclass = {quant-ph},
	title = {{Entanglement entropy production in deep inelastic scattering}},
	volume = {105},
	year = {2022}}

@article{Baker:2017wtt,
	archiveprefix = {arXiv},
	author = {Baker, O. K. and Kharzeev, D. E.},
	doi = {10.1103/PhysRevD.98.054007},
	eprint = {1712.04558},
	journal = {Phys. Rev. D},
	number = {5},
	pages = {054007},
	primaryclass = {hep-ph},
	title = {{Thermal radiation and entanglement in proton-proton collisions at energies available at the CERN Large Hadron Collider}},
	volume = {98},
	year = {2018}}

@article{Kharzeev:2021yyf,
	archiveprefix = {arXiv},
	author = {Kharzeev, Dmitri E. and Levin, Eugene},
	doi = {10.1103/PhysRevD.104.L031503},
	eprint = {2102.09773},
	journal = {Phys. Rev. D},
	number = {3},
	pages = {L031503},
	primaryclass = {hep-ph},
	title = {{Deep inelastic scattering as a probe of entanglement: Confronting experimental data}},
	volume = {104},
	year = {2021}}

@article{Kharzeev:2017qzs,
	archiveprefix = {arXiv},
	author = {Kharzeev, Dmitri E. and Levin, Eugene M.},
	doi = {10.1103/PhysRevD.95.114008},
	eprint = {1702.03489},
	journal = {Phys. Rev. D},
	number = {11},
	pages = {114008},
	primaryclass = {hep-ph},
	title = {{Deep inelastic scattering as a probe of entanglement}},
	volume = {95},
	year = {2017}}

@article{Fries:2008vp,
	archiveprefix = {arXiv},
	author = {Fries, Rainer J. and M\"uller, Berndt and Sch\"afer, Andreas},
	doi = {10.1103/PhysRevC.79.034904},
	eprint = {0807.1093},
	journal = {Phys. Rev. C},
	pages = {034904},
	primaryclass = {nucl-th},
	title = {{Decoherence and Entropy Production in Relativistic Nuclear Collisions}},
	volume = {79},
	year = {2009}}

@article{Kovner:1995ja,
	archiveprefix = {arXiv},
	author = {Kovner, Alex and McLerran, Larry D. and Weigert, Heribert},
	doi = {10.1103/PhysRevD.52.6231},
	eprint = {hep-ph/9502289},
	journal = {Phys. Rev. D},
	pages = {6231--6237},
	reportnumber = {TPI-MINN-95-02-T, NUC-MINN-95-8-T, HEP-MINN-95-1327, TPI--MINN--95--02-T, NUC--MINN--95--8-T, HEP--MINN--95--1327},
	title = {{Gluon production from nonAbelian Weizsacker-Williams fields in nucleus-nucleus collisions}},
	volume = {52},
	year = {1995}}

@article{Becattini:2008tx,
	archiveprefix = {arXiv},
	author = {Becattini, F. and Castorina, P. and Manninen, J. and Satz, H.},
	doi = {10.1140/epjc/s10052-008-0671-x},
	eprint = {0805.0964},
	journal = {Eur. Phys. J. C},
	pages = {493--510},
	primaryclass = {hep-ph},
	title = {{The Thermal Production of Strange and Non-Strange Hadrons in e+ e- Collisions}},
	volume = {56},
	year = {2008}}

@article{Cleymans:1992zc,
	archiveprefix = {arXiv},
	author = {Cleymans, J. and Satz, H.},
	doi = {10.1007/BF01555746},
	eprint = {hep-ph/9207204},
	journal = {Z. Phys. C},
	pages = {135--148},
	reportnumber = {BI-TP-92-08, CERN-TH-6523-92},
	title = {{Thermal hadron production in high-energy heavy ion collisions}},
	volume = {57},
	year = {1993}}

@article{Fermi:1950jd,
	author = {Fermi, Enrico},
	doi = {10.1143/PTP.5.570},
	journal = {Prog. Theor. Phys.},
	pages = {570--583},
	title = {{High-energy nuclear events}},
	volume = {5},
	year = {1950}}

@article{Chen:2026hnh,
    author = {Chen, Vincent and M{\"u}ller, Berndt and Yao, Xiaojun},
    title = "{Minimally Truncated SU(3) Lattice Gauge Theory and String Tension}",
    eprint = "2601.10065",
    archivePrefix = "arXiv",
    primaryClass = "hep-lat",
    reportNumber = "IQuS@UW-21-118",
    month = "1",
    year = "2026"
}

@article{Grieninger:2026bdq,
    author = "Grieninger, Sebastian and Savage, Martin J. and Zemlevskiy, Nikita A.",
    title = "{The Quantum Complexity of String Breaking in the Schwinger Model}",
    eprint = "2601.08825",
    archivePrefix = "arXiv",
    primaryClass = "hep-ph",
    reportNumber = "IQuS@UW-21-119, NT@UW-26-1",
    month = "1",
    year = "2026"
}

@article{Liu:2024lut,
    author = "Liu, Ying and Zhang, Wei-Yong and Zhu, Zi-Hang and He, Ming-Gen and Yuan, Zhen-Sheng and Pan, Jian-Wei",
    title = "{String-Breaking Mechanism in a Lattice Schwinger Model Simulator}",
    eprint = "2411.15443",
    archivePrefix = "arXiv",
    primaryClass = "cond-mat.quant-gas",
    doi = "10.1103/mwy1-v9hk",
    journal = "Phys. Rev. Lett.",
    volume = "135",
    number = "10",
    pages = "101902",
    year = "2025"
}

@article{Lee:2023urk,
    author = "Lee, Kyle and Mulligan, James and Ringer, Felix and Yao, Xiaojun",
    title = "{Liouvillian dynamics of the open Schwinger model: String breaking and kinetic dissipation in a thermal medium}",
    eprint = "2308.03878",
    archivePrefix = "arXiv",
    primaryClass = "quant-ph",
    reportNumber = "JLAB-THY-23-3894, MIT-CTP-5592, YITP-SB-2023-23, IQuS@UW-21-061",
    doi = "10.1103/PhysRevD.108.094518",
    journal = "Phys. Rev. D",
    volume = "108",
    number = "9",
    pages = "094518",
    year = "2023"
}

@article{Ikeda:2025bjb,
    author = "Ikeda, Kazuki and Kang, Zhong-Bo and Kharzeev, Dmitri E. and Qian, Wenyang",
    title = "{Quantum simulation of deep inelastic scattering in the Schwinger model}",
    eprint = "2512.18062",
    archivePrefix = "arXiv",
    primaryClass = "hep-ph",
    month = "12",
    year = "2025"
}

@article{Florio:2025gic,
    author = "Florio, Adrien and Frenklakh, David and Ikeda, Kazuki and Kharzeev, Dmitri and Korepin, Vladimir and Shi, Shuzhe and Yu, Kwangmin",
    title = "{Quantum simulation of entanglement and hadronization in high-energy collisions: lessons from the massive Schwinger model}",
    doi = "10.22323/1.483.0056",
    journal = "PoS",
    volume = "QCHSC24",
    pages = "056",
    year = "2025"
}

@article{Florio:2025hoc,
    author = "Florio, Adrien and Frenklakh, David and Grieninger, Sebastian and Kharzeev, Dmitri E. and Palermo, Andrea and Shi, Shuzhe",
    title = "{Thermalization from quantum entanglement: Jet simulations in the massive Schwinger model}",
    eprint = "2506.14983",
    archivePrefix = "arXiv",
    primaryClass = "hep-ph",
    doi = "10.1103/sgrx-jpp9",
    journal = "Phys. Rev. D",
    volume = "112",
    number = "9",
    pages = "094502",
    year = "2025"
}

@article{Florio:2024aix,
    author = "Florio, Adrien and Frenklakh, David and Ikeda, Kazuki and Kharzeev, Dmitri E. and Korepin, Vladimir and Shi, Shuzhe and Yu, Kwangmin",
    title = "{Quantum real-time evolution of entanglement and hadronization in jet production: Lessons from the massive Schwinger model}",
    eprint = "2404.00087",
    archivePrefix = "arXiv",
    primaryClass = "hep-ph",
    doi = "10.1103/PhysRevD.110.094029",
    journal = "Phys. Rev. D",
    volume = "110",
    number = "9",
    pages = "094029",
    year = "2024"
}

@article{Grieninger:2023ufa,
    author = "Grieninger, Sebastian and Ikeda, Kazuki and Kharzeev, Dmitri E. and Zahed, Ismail",
    title = "{Entanglement in massive Schwinger model at finite temperature and density}",
    eprint = "2312.03172",
    archivePrefix = "arXiv",
    primaryClass = "hep-th",
    doi = "10.1103/PhysRevD.109.016023",
    journal = "Phys. Rev. D",
    volume = "109",
    number = "1",
    pages = "016023",
    year = "2024"
}

@article{Florio:2023dke,
    author = "Florio, Adrien and Frenklakh, David and Ikeda, Kazuki and Kharzeev, Dmitri and Korepin, Vladimir and Shi, Shuzhe and Yu, Kwangmin",
    title = "{Real-Time Nonperturbative Dynamics of Jet Production in Schwinger Model: Quantum Entanglement and Vacuum Modification}",
    eprint = "2301.11991",
    archivePrefix = "arXiv",
    primaryClass = "hep-ph",
    doi = "10.1103/PhysRevLett.131.021902",
    journal = "Phys. Rev. Lett.",
    volume = "131",
    number = "2",
    pages = "021902",
    year = "2023"
}

@article{Buividovich:2008kq,
    author = "Buividovich, P. V. and Polikarpov, M. I.",
    title = "{Numerical study of entanglement entropy in SU(2) lattice gauge theory}",
    eprint = "0802.4247",
    archivePrefix = "arXiv",
    primaryClass = "hep-lat",
    reportNumber = "ITEP-LAT-2008-07",
    doi = "10.1016/j.nuclphysb.2008.04.024",
    journal = "Nucl. Phys. B",
    volume = "802",
    pages = "458--474",
    year = "2008"
}

@article{Itou:2015cyu,
    author = "Itou, Etsuko and Nagata, Keitaro and Nakagawa, Yoshiyuki and Nakamura, Atsushi and Zakharov, V. I.",
    title = "{Entanglement in Four-Dimensional SU(3) Gauge Theory}",
    eprint = "1512.01334",
    archivePrefix = "arXiv",
    primaryClass = "hep-th",
    reportNumber = "KEK-CP-337",
    doi = "10.1093/ptep/ptw050",
    journal = "PTEP",
    volume = "2016",
    number = "6",
    pages = "061B01",
    year = "2016"
}

@article{Rabenstein:2018bri,
    author = {Rabenstein, Andreas and Bodendorfer, Norbert and Buividovich, Pavel and Sch{\"a}fer, Andreas},
    title = "{Lattice study of R{\'e}nyi entanglement entropy in $SU(N_c)$ lattice Yang-Mills theory with $N_c = 2, 3, 4$}",
    eprint = "1812.04279",
    archivePrefix = "arXiv",
    primaryClass = "hep-lat",
    doi = "10.1103/PhysRevD.100.034504",
    journal = "Phys. Rev. D",
    volume = "100",
    number = "3",
    pages = "034504",
    year = "2019"
}

@article{Fujii:2025aip,
    author = "Fujii, Daisuke and Kawaguchi, Mamiya and Tanaka, Mitsuru",
    title = "{Dominance of gluonic scale anomaly in confining pressure inside nucleon and D-term}",
    eprint = "2503.09686",
    archivePrefix = "arXiv",
    primaryClass = "hep-ph",
    doi = "10.1016/j.physletb.2025.139559",
    journal = "Phys. Lett. B",
    volume = "866",
    pages = "139559",
    year = "2025"
}

@article{Burkert:2018bqq,
    author = "Burkert, V. D. and Elouadrhiri, L. and Girod, F. X.",
    title = "{The pressure distribution inside the proton}",
    doi = "10.1038/s41586-018-0060-z",
    journal = "Nature",
    volume = "557",
    number = "7705",
    pages = "396--399",
    year = "2018"
}

@article{Bissey:2005sk,
    author = "Bissey, F. and Cao, F-G. and Kitson, A. and Lasscock, B. G. and Leinweber, D. B. and Signal, A. I. and Williams, A. G. and Zanotti, J. M.",
    editor = "Kizilersu, Ayse and Williams, Anthony G. and Thomas, Anthony W.",
    title = "{Gluon field distribution in baryons}",
    eprint = "hep-lat/0501004",
    archivePrefix = "arXiv",
    doi = "10.1016/j.nuclphysbps.2004.12.004",
    journal = "Nucl. Phys. B Proc. Suppl.",
    volume = "141",
    pages = "22--25",
    year = "2005"
}

\end{document}